\pdfoutput=1

\documentclass[11pt]{article}
\usepackage{jheppub}

\usepackage{amsmath}
\usepackage{latexsym,amssymb}
\usepackage{graphicx,color,slashed}
\usepackage{bbm} 
\usepackage{mathtools}
\usepackage{ifpdf}

\newcommand{\cG}{\mathcal{G}}

\newcommand{\cH}{\mathcal{H}}

\newcommand{\cN}{\mathcal{N}}

\newcommand{\be}{\begin{equation}}
\newcommand{\ee}{\end{equation}}
\newcommand{\ba}{\begin{eqnarray}}
\newcommand{\ea}{\end{eqnarray}}

\renewcommand{\a}{\alpha}

\def\E{{$E_{7(7)}$}}

\newcommand{\rf}[1]{(\ref{#1})}
\newcommand{\bea}{\begin{eqnarray}}
\newcommand{\eea}{\end{eqnarray}}

\def\bfzero{\relax{\rm I\kern-.18em 0}}
\def\bfone{\relax{\rm 1\kern-.35em 1}}
\def\twomat#1#2#3#4{\left(\begin{array}{cc}
\end{array}
\right)}

\def \CC {{\rm C}}

\newcommand{\ui} {\underline i} 

\newcommand{\uj} {\underline j}

\newcommand{\nonu}{\nonumber \\[.5mm]}
\newcommand{\A}{&\!\!\!}

\title{\rm{\bf     Ward Identities for Superamplitudes}}
\author
{ Renata Kallosh}
 \affiliation{Stanford Institute for Theoretical Physics and Department of Physics,\\
  Stanford University, Stanford, CA 94305, USA}

\abstract{  We introduce Ward identities for superamplitudes in D-dimensional  $\cN$-extended supergravities. These identities help to clarify the relation between linearized superinvariants and superamplitudes. The solutions of these Ward identities for an $n$-partice superamplitude take a simple universal form for half  BPS and non-BPS amplitudes. These solutions involve arbitrary functions of spinor helicity and Grassmann variables for each of the $n$ superparticles. The dimension of these functions at a given loop order is exactly the same as the dimension of the relevant superspace Lagrangians depending on half-BPS or non-BPS superfields, given by $(D-2) L +2- \cN$ or $(D-2) L +2- 2\cN$, respectively. This explains why soft limits predictions from superamplitudes and from superspace linearized superinvariants agree. 
}

\begin{document}

\maketitle



\parskip 5pt


\section{Introduction}

Supersymmetry and R-symmetry Ward identities impose linear relations among $n$-point  on-shell supergravity amplitudes.  The superamplitudes satisfying Ward  identities are consistent with linearized supersymmetry.

It is also possible to study linearized supersymmetry constraints in supergravity using superinvariants. These are integrals over linearized superspace of Lagrangians depending on linearized superfiedlds. In this case, a dimensional analysis is available, and it leads to specific predictions with regard to candidate counterterms and soft scalar limits.

A relation between superamplitudes and linearized superinvariants is known in various examples. Here, we will present a universal relation that exists between these two approaches to linearized supersymmetry in various dimensions.

In 4D supergravities, Ward identities for multi-point superamplitudes are known in maximal supergravity case \cite{Bianchi:2008pu}. 
Explicit solutions to these Ward identities were given in \cite{Elvang:2009wd}. The method developed in 
 \cite{Elvang:2009wd} to produce superamplitudes, solutions of Ward identities for $\cN=8$, was adapted to 4D supergravities with  $\cN=5,6$ in \cite{Freedman:2018mrv}. The purpose of studying 6-point superamplitudes in \cite{Freedman:2018mrv} was to understand their soft scalar behavior. It was also observed in \cite{Freedman:2018mrv} that the study of local superinvariants gave the same results as the actual solutions of Ward identities for 6-point superamplitudes in all cases $\cN\geq 5$.
 
 Ward identities for superamplitudes were explored in general for higher dimensions; see, for example, \cite{Boels:2012ie} and references therein. In type IIB 10D supergravity in  \cite{Wang:2015jna,Green:2019rhz} the form of the Ward identities for the amplitudes in maximal 10D supergravity was proposed and investigated. Here  we will generalize it to $D\geq 4$ $\cN$-extended supergravities.  
 
Recently, a set of local 6-point superinvariants was studied in \cite{Kallosh:2023dpr} in 6D maximal supergravity. It was observed\footnote{C. Wen, Y.-t. Huang, private communication and  \cite{Kallosh:2023dpr}.} that the 
properties of superamplitudes in 10D and superinvariants in 6D are consistent.
  
  The purpose of this note is to propose a form of Ward identities for multi-point superamplitudes for supergravities in diverse integer dimensions. In particular, this will make clear the relation between linearized superinvariants and superamplitudes for D-dimensional supergravities with $\cN$-extended supersymmetries.

Our main applications of the studies of Ward identities for amplitudes will be with regard to cases where there is no 1-loop anomaly in R-symmetry. Namely, in 4D, there are no 1-loop R-symmetry anomalies at $\cN\geq 5$, and in 6D, there is no 1-loop R-symmetry anomaly in maximal supergravity \cite{Marcus:1985yy}. But in both cases, the status of soft scalar limits of the 6-point amplitudes, i.e., the status of 1-loop $\cG$-symmetry anomalies,  has not been established.

We will find out that the solution of the Ward identities for local amplitudes combined with dimensional analysis directly gives correct results concerning soft scalar limits without the need to perform complicated calculations of the explicit 6-point superamplitudes.

\section{Superamplitudes and Ward Identities in D-dimensional supergravity}

Consider multi-point superamplitudes, which are constrained by the requirement of linearized supersymmetry. Let us first ignore the issue of dimensions, loop order, and power of gravitational coupling. We just require the $n$-point superamplitudes to be linearly supersymmetric, i.e., satisfy Ward identities. The nice form of on-shell supersymmetry Ward identities for amplitudes in 10D type IIB maximal supergravity was given in  \cite{Wang:2015jna}. We generalize it here for other dimensions and $\cN$-extended supergravities. 

Following \cite{Tanii:1998px} we describe the 
field contents of ungauged supergravities as determined by irreducible 
representations of the super Poincar\'e algebras. 
There are
 translations $P_a$, 
generators of Lorentz transformations $M_{ab}$, 
supercharges $Q^{\alpha \ui}$, and generators of automorphism group $T^A$. In cases of massless states, there are no central charges. 
\be
[ M_{ab}, Q^{\ui }]  =   {1 \over 2} \gamma_{ab} Q^{\ui }\ , \qquad
[ T^A, Q^{\ui } ] = (t^A)^{\ui }{}_{\uj}Q^{\uj}\ , \qquad
[ T^A, T^B ] = f^{AB}{}_C T^C \ . 
\ee
Here
 $t^A$ and $f^{AB}{}_C$ are representation matrices and 
 structure constant of the Lie algebra of the automorphism group.
The automorphism group is R, and the form of anticommutators 
$\{ Q, Q \}$ depends on the spinor type of $Q^{\ui }$. 

The spinors in various dimensions have 
different properties described in  \cite{Tanii:1998px}. They could be
Weyl, Majorana, symplectic (pseudo) Majorana, Majorana-Weyl, and symplectic (pseudo) Majorana-Weyl spinors. In even dimensions, the supersymmetry algebra can be presented via spinors of positive and negative chirality. In odd dimensions, the supercharges are (pseudo) Majorana in 9D and 11D, and symplectic (pseudo) Majorana in 5D and 7D.

In all dimensions, there is a choice of supersymmetries with a total amount of $4\cN $, which can be split in half,  $2\cN $+$2\cN $.  The choice in each case is defined by the choice of the 1/2 BPS superfields, which depend on 1/2 of fermionic directions. It is known in each dimension. In applications to $n$-point amplitudes we define these two sets as $Q^A$ and $\tilde Q^A$
\be
Q^A = \sum_{i=1}^n q^A_i = \sum_{i=1}^n (\lambda_i \, \eta_i)^A\ ,  \quad \tilde Q^A = \sum_{i=1}^n \tilde q^A_i = \sum_{i=1}^n \Big (\lambda_i {\partial \over \partial \eta_i }\Big)^A \ , \quad i=1\ ,\dots , n \ , \quad A=1,\dots , 2\cN \ , \nonumber\\
\ee
 where
 \be
\{ Q^A, {\tilde Q}^B \}= \gamma^{AB}_a P^a\, ,  \qquad \{ Q^A,  Q^B \}=0\, ,  \qquad \{ {\tilde Q}^A, {\tilde Q}^B \}=0 \ .
\label{susy}\ee
The $n$-point amplitude is defined as 
\be
{\cal A}=\delta^D(P)\delta^{2\cN}(Q) {\cal F} (\lambda_i, \eta_i) \ .
\ee
The amplitude is manifestly supersymmetric under the action of the $Q$ operator, i. e. under 1/2 of supersymmetry. However, the action of the operator $\tilde Q$ leads to Ward identities
\be\boxed{
\tilde Q {\cal A}=\delta^D(P)\delta^{2\cN }(Q) \tilde Q [ {\cal F} (\lambda_i, \eta_i)]=0}
\label{WI}\ee 
There are 3 types of solutions given in \cite{Wang:2015jna} in the 10D case. These are also valid in general dimensions\footnote{See however Appendix \ref{App:A} here where we discuss one-quarter and one-eighth linearised BPS superinvariants.}.

1. The first solution requires that   $ {\cal F} 
$ is $\eta$ independent
\be
 {\cal F} (\lambda_i, \eta_i)=  {\cal F} (\lambda_i)= f(s_{ij}) \ .
\ee
The supersymmetric amplitude is
\be
{\cal A}=\delta^D(P)\delta^{2\cN}(Q) f(s_{ij})
\label{simple}\ee
since $\{ Q, {\tilde Q} \}= \gamma_a P^a\, ,$ and there is a delta-function $\delta^D(P)$, the amplitude has all $4\cN$ supersymmetries. These are superamplitudes of the 1/2 BPS type.

2. A CPT conjugate  amplitude ${\overline A}$ is also supersymmetric. An example in the 10D case is given in eq. (6) of  \cite{Wang:2015jna}.

3. The function $ {\cal F} 
$ is $\eta$-dependent, and this dependence is strongly constrained by the Ward Identity. The function $ {\cal F} (\lambda_i, \eta_i)$ has to be of the form 
\be
{\cal F} (\lambda_i, \eta_i)=\tilde Q \, h(\lambda_i, \eta_i)=  \sum_{j=1}^n \lambda_j \, {\partial \over \partial \eta_j } h(\lambda_i, \eta_i) 
\ee
and Ward identities \rf{WI} are satisfied
\be
\tilde Q [ {\cal F} (\lambda_i, \eta_i)] = \tilde Q^2 \, h(\lambda_i, \eta_i)=0\, .
\ee
Therefore the  solution to Ward identities is 
\be
\tilde Q \, {\cal A}= 0 \quad \rightarrow \quad  {\cal A} = \delta^D(P)\delta^{2\cN}(Q) \tilde Q \, h(\lambda_i, \eta_i) \ .
\label{WIS}\ee
Thus, the $n$-point amplitude satisfying supersymmetric Ward identities has the form given in \rf{WIS} where the function $h(\lambda_i, \eta_i)$ in not constrained as long as the total amplitude has a factor 
\be
\delta^{2\cN}(Q) \tilde Q \ ,
\ee
which has dimension $2\cN$ as this expression involves the product of $4\cN$ $\lambda$'s which are  spinor helicity variables of dimension 1/2. Note that in amplitudes, the $\eta$'s are dimensionless variables. Thus we observe that these superamplitudes are associated with superinvariants where the $\theta$ and $\tilde \theta$ variables have dimension 1/2.
\be
\delta^{2\cN}(Q) \tilde Q \qquad \Longleftrightarrow \qquad  d^{2\cN} \theta d^{2\cN} \tilde \theta \ , \qquad {\rm dim} \, 2\cN \ .
\label{whole}\ee
These superamplitudes are non-BPS type.

\section{Dimensional analysis of superamplitudes}
Consider the amplitudes at the $L$-loop order in D-dimensional $\cN$-extended supergravity. In front of the L-loop superamplitude, we have a gravitational coupling 
\be
\kappa^{2(L-1)} {\cal A} 
\label{dim}\ee
where dimension of  $\kappa^2$ is $(2-D)$.  

1. The half-BPS amplitudes are
\be
\kappa^{2(L-1)} {\cal A}_{BPS}= \kappa^{2(L-1)} \delta^D(P)\delta^{2\cN}(Q) f(s_{ij}) \ .
\ee
 therefore the dimension counting is $(2-D)(L-1) - D + \cN + {\rm dim} f(s_{ij})=0$ and 
\be
{\rm dim}[ f(\lambda_i)]= (D-2) L +2-\cN \ .
\label{f}\ee
Consider an example of 6D maximal $\cN=8$ supergravity 4-point superamplitude. At $L=0$ $ {\rm dim} [f(s_{ij})]= -6$, at $L=3$   ${\rm dim} [f(s_{ij})]= 6$. This agree with the $L=0$ result in \cite{Cachazo:2018hqa} where $f={1\over stu}$ and   $L=3$ result in \cite{Kallosh:2023dpr} where $f= stu$.

Another example is 4D maximal supergravity 4-point amplitudes. At $L=0$ $ {\rm dim} [f(s_{ij})]= -6$, at $L=3$   ${\rm dim} [f(s_{ij})]=0$. This agree with the $L=0$ result in \cite{Kallosh:2008ru} where $f={1\over stu}$ and   $L=3$ where $f= const$.

2. The non-BPS $n$-point amplitudes are
\be
\kappa^{2(L-1)} {\cal A}_{non-BPS} = \kappa^{2(L-1)} \delta^D(P)\delta^{2\cN}(Q) \tilde Q \, h(\lambda_i, \eta_i) \ .
\label{nBPS}\ee
Dimension counting is $(2-D)(L-1) - D + 2\cN + {\rm dim} [ h(\lambda_i, \eta_i)]=0$ and 
\be
{\rm dim} [ h(\lambda_i, \eta_i)]= (D-2) L +2- 2\cN \ .
\label{nBPSc}\ee
We find that to have supervertices, i.e., superamplitudes with no poles in momenta with ${\rm dim} [ h(\lambda_i, \eta_i)]\geq 0$ requires that
\be
(D-2) L +2- 2\cN \geq 0 \ .
\label{nBPSloc}\ee
For example, in 4D we need 
$
 L  \geq  \cN -1
$
 to have supervertices (no poles). But at $L\leq \cN -2$ these supervertices are not available, which results in a non-vanishing scalar soft limits and broken \E\, symmetry if there were UV divergences at $L\leq \cN-2$.
This is in full agreement with \cite{Beisert:2010jx} for $\cN=8$ and with \cite{Freedman:2018mrv} for $\cN\geq 5$. Moreover, we have observed that already based on supersymmetric Ward identities and dimensional analysis of superamplitudes, it was possible to establish this result without explicit computations of the 6-point superamplitudes. 

\section{Dimensional analysis of linearized superinvariants} 
Half-BPS linearized superinvariants are
\be
S^{BPS}= \kappa^{2(L-1)} \int d^{2\cN} \theta d^D x \, {\cal L}^{BPS} \ .
\ee
Dimension of $\kappa^2$ is $2-D$. For the action to be dimensionless we require that 
\be
 {\rm dim} \,{\cal L}^{BPS}= L(D-2)+2   - \cN 
\label{rulesBPS}\ee
For non-BPS superinvariants we have
\be
S^{non-BPS}= \kappa^{2(L-1)} \int d^{4\cN} \theta d^D x \, {\cal L}^{non-BPS} \ .
\label{nBPS}\ee
and
\be
{\rm dim} \,{\cal L}^{non-BPS}= L(D-2)+2   - 2\cN 
\label{rules}\ee
In amplitude applications, what we would like to call ``local multi-point superinvariants''  are often called supervertices, namely superamplitudes with no poles in momenta, see, for example, \cite{Wang:2015jna}. 
Let us find out the smallest loop order where $n$-point supervertices are available. Consider superinvariants in eq. \rf{nBPS}
where
\be
L(2-D)-2  + 2\cN + {\rm dim} \,{\cal L}^{non-BPS}= 0  \ ,
\label{rules}\ee
and 
\be
L=  {2(\cN-1) + {\rm dim} \,{\cal L}^{non-BPS}\over (D-2)} \ .
\ee
Below this loop order, there are no local multi-point superinvariants, i.e., there are no superamplitudes without the poles in momenta.

Consider known examples where for local superinvariants we require that
\be
{\rm dim} \,{\cal L} \geq 0 \ .
\ee
In 4D we can take $
 {\rm dim} \,{\cal L}=0
$   as it depends on a product of $n$ superfields with scalars in the first component. The R-invariant $n$-point local multi-point superinvariants with $n\geq 6$ according to \rf{rules} are available starting with 
\be
L= \cN-1\, , \qquad D=4 \ ,
\ee
and they are not available at $L\leq  \cN-2$ in 4D. This is a result  obtained in
 \cite{Freedman:2018mrv}  based on analysis of superinvariants.  
 
This simple statement for superinvariants was confirmed by direct computations of N$^k$MHV 6-point superamplitudes in $\cN=5,6,8$ supergravities  \cite{Freedman:2018mrv}. The result of the computation, using solutions of Ward identities, did confirm the linearized superinvariant analysis.

Now we take 6D maximal supergravity where there is a UV divergence at $L=3$ \cite{Bern:2008pv}. It was found in \cite{Kallosh:2023dpr}    that local multi-point superinvariants R-symmetry invariants are available starting with 
\be
L={14+ {\rm dim} \,{\cal L}\over 4}  > 3\, ,  \qquad D=6\, ,  \qquad \cN=8  \ .
\ee
But unlike in the 4D case, the computation of the relevant 6-point superamplitude has not yet been performed. It would confirm that at L=3 there are no R-invariant local (no poles) superamplitudes.

In 8D, one finds  in maximal supergravity case where there is a UV divergence at $L=1$ \cite{Bern:1998ug}  that local multi-point superinvariants are available starting with 
\be
L={14 +{\rm dim} \,  {\cal L}\over 6} > 2 \, , \qquad D=8 \, , \qquad \cN=8  \ .
\ee

\section{R symmetry}

The supervertices in eq. \rf{simple}, which we call half-BPS amplitudes in all dimensions,  are also called F-term vertices in 10D case  \cite{Wang:2015jna}, and MRV (Maximal R-symmetry Violating) amplitudes in \cite{Boels:2012zr}.
In 4D, these are known as MHV, Maximally Helicity Violating amplitudes, but these do not exist in other dimensions \cite{Boels:2012zr}.  

4-point superamplitudes/superinvariants make an exception in all dimensions. The BPS 4-point amplitudes do not break R-symmetry.
The existence of BPS amplitudes consistent with R-symmetry for 4-point amplitudes is known in 4D supergravities \cite{Kallosh:1980fi,Howe:1981xy} where it was proven for local BPS superinvariants. In 10D, it was proven in \cite{Boels:2012zr} that for $n>4$, BPS superamplitudes R-symmetry is broken, but it is preserved for $n=4$. Finally, 
 in 6D, it was proven in \cite{Kallosh:2023dpr} that 4-point BPS superinvariants have unbroken R-symmetry, but the ones with $n>4$ break R-symmetry.

Also, in known examples, the 4-point superamplitudes defined as half-BPS invariants in eq. \rf{simple} are consistent with R-symmetry. However, at higher point amplitudes $n>4$, the issue of R-symmetry is more complicated, and these BPS superamplitudes often break R-symmetry. 

However, there are cases of $n$-point half-BPS superamplitudes/linearized superinvariants with $n>4$, which preserve R-symmetry. For example, in  \cite{Huang:2015sla} in 4D $\cN=4$ supergravity, there is a half-BPS 5-point amplitude which is not anomalous under $U(1)$ symmetry. It corresponds to a rational part of the 5-point amplitude $R_5$ computed in \cite{Dunbar:2010fy}. The corresponding $U(1)$ preserving superamplitude is given in eq. (4.13) in  \cite{Huang:2015sla}. Here, we will present it in the form of the linearized superinvariant
\be
{\rm MHV} ^{(3,2)}= M^1+M^2
\ee
where
\be
M^1 =  \int d^8 \bar \theta \, 
{1 \over s_{45}s_{43}s_{53}}\overline W_1  \overline W_2 C_3 *C_4 * C_5  
\ee
 and 
\be
M^2= \int d^8 \bar \theta \, {1\over s_{34}^2 s_{25} s_{35} s_{45}} {\overline W}_1\, \partial^\alpha{}_{\dot \alpha} \partial^\beta{}_{\dot \beta } {\overline W}_2 (  \partial^\gamma{}_{\dot \alpha } C_{ 3 } *\partial^\delta{}_{\dot \beta } \, C_4  ) C_{ 5\,  \alpha \beta \gamma\delta} + \{1,2\} + \{3,4,5\}
\ee
Here
\be
\overline W_i\equiv \overline W (p_i, \bar \theta)\, ,   \qquad C_3 *C_4 * C_5  \equiv  \CC_{\alpha\beta}{}^{\gamma\delta} (p_3, \bar \theta) \CC_{\gamma\delta}{}^{\mu\nu} (p_4, \bar \theta)  \CC_{\mu\nu}{}^{\alpha\beta} (p_5, \bar \theta) 
\ee
\be
\partial  C_3 *\partial C_4   \equiv \partial C_{\alpha\beta}{}^{\gamma\delta} (p_3, \bar \theta) \partial\CC_{\gamma\delta}{}^{\alpha\beta} (p_4, \bar \theta)  
\ee
The fact that this half-BPS 5-point superinvariants/superamplitudes preserve $U(1)$ is obvious. The Weyl superfields $\CC_{\alpha\beta}{}^{\gamma\delta}$ have zero $U(1)$ charge, and two $\bar W$' have  $U(1)$ charge canceling the ones from a half-BPS measure of integration. 

In these superinvariants 
\be
{\rm dim} [{\cal L}^1(x, \bar \theta)]=  {\rm dim} [{1 \over \partial^6}\overline W^2 C^3]=0\, \qquad {\rm dim} [ {\cal L}^2(x, \bar \theta)] ={\rm dim} [ {1\over \partial^{10}} {\overline W}\, \partial^2 {\overline W}   \partial C \partial C C]=0
\ee
We can compare it with explicit expressions for the superamplitudes  for  \cite{Huang:2015sla} in half-BPS \rf{simple}
\be
{\rm dim} [f^1(\lambda_i)] ={\rm dim} [  {[34][45][53]\over \langle 34\rangle  \langle 45\rangle  \langle 53 \rangle} ]=0\, , \qquad {\rm dim} [f^2(\lambda_i)] ={\rm dim} [{[34]^2\over \langle 34\rangle^2}  {[25]\langle 23\rangle \langle 24\rangle\over \langle 25\rangle  \langle 35\rangle  \langle 45 \rangle} ]=0
\ee
This is an example of the R-invariant half-BPS 5-point superamplitude in eq. \rf{f} in 4D at $\cN=4$ and $L=1$
\be
{\rm dim}[ f(\lambda_i)]= (D-2) L +2-\cN = 2+2-4=0
\label{fWY}\ee

On the other hand, the known examples of non-BPS $n$-point amplitudes with $n>4$ satisfy Ward identities in eq. \rf{WIS} can easily preserve R-symmetry. One can see from eq. \rf {whole} that the part of the superamplitude has a factor $\delta^{2\cN}(Q) \tilde Q$ analogous to the measure of the integration over the whole superspace $d^{2\cN} \theta d^{2\cN} \tilde \theta$ which has R-symmetry. 

To keep R-symmetry using whole superspace superinvariants, one needs to have also a Lagrangian  $\cal L$ preserving R-symmetry.
\be
 \int d^{2\cN} \theta d^{2\cN} \tilde \theta d^D x \, {\cal L}(x, \theta, \tilde \theta) \ .
\ee
In superamplitudes \rf{WIS}, the function $h(\lambda_i, \eta_i)$ has to be taken in the form preserving R-symmetry to have R-symmetric $n$-point superamplitudes.

\section{Application to 4D $\cN\geq 5$ supergravity soft limits }\label{Sec:4d}

In   \cite{Freedman:2018mrv}, the computation of 6-point superamplitudes was performed. The results were that the local superamplitudes are available starting from $L=\cN-1$ but not below this loop order.

Here, instead of computing 6-point superamplitudes satisfying Ward identities, we present an 
alternative derivation of the same result. R-symmetry invariant $n$-point amplitudes with $n>4$ require the form of solutions of Ward identities given in \rf{nBPS},  \rf{nBPSc}, \rf{nBPSloc}

Supervertices (superamplitudes without poles) are available for $n$-point amplitudes, $n>4$ in  4D supergravity starting with 
\be
{\rm dim} [ h(\lambda_i, \eta_i)]= 2 L +2- 2\cN= 2(L-(\cN-1))\geq 0 \ .
\label{dimh}\ee
This means that  6-point superamplitudes, supervertices without poles,  are available at 
\be
L\geq \cN-1 \ .
\ee
This confirms the computation of 6-point superamplitudes in \cite{Freedman:2018mrv} where it was found that all constructed 6-point superamplitudes with non-vanishing soft scalar limit are not local.

Here, we have found a way around the complicated computation of all explicit 6-point amplitudes of the kind performed in \cite{Freedman:2018mrv}. It was possible to use a non-explicit solution of 4D Ward identities in eq. \rf{nBPS} and only study the dimension of the relevant function $h(\lambda_i, \eta_i)$.

Let us illustrate the case with $\cN=8, L=6$ where there are no local 6-point superamplitudes, according to \rf{dimh}.

To study soft limits, we need a 6-point 
 superinvariant with bosonic part
\be
\kappa^{10} \int d^4x\,  \partial^6 \, R^4 \, \phi_{ijkl} \, \bar \phi^{ijkl} \ .
\ee
We can try
\be
\kappa^{10} \, \int \, d^4x \, d^{32} \theta \, \partial^{2l } (W_{ijkl} \, \bar W^{ijkl})^3 \ .
\ee
It means that 
 $-10-4 +16 +2l=0$, and $l=-1$ so the relevant supersymmetric expression is
\be
\kappa^{10} \, \int \, d^4x \, d^{32} \theta \, (W_{ijkl} \, \bar W^{ijkl})^3 \partial^{-2} \ ,
\ee
and it is non-local. Using superamplitudes \rf{dimh} we find the same dimension of $h$ as dimension of ${\cal L}=\partial^{2l } (W_{ijkl} \, \bar W^{ijkl})^3$ with $l=-1$
\be
{\rm dim} [ h(\lambda_i, \eta_i)]=  2(6-7)= -2 \, .
\ee
Thus, we see that at $L=6$, maximal 4D supergravity does not have a local expression with bosonic part $ \partial^6 \, R^4 \, \phi \, \bar \phi$. This, in turn, means that the 4-point UV divergence at $L=6$ maximal 4D supergravity would imply that the \E\ symmetry is broken. This conclusion was reached in \cite{Beisert:2010jx} using string theory and in \cite{Freedman:2018mrv} using superinvariants. Here we 
deduced the same result using dimensional analysis of 4D superamplitudes satisfying Ward identities.

\section{Applications to 6D maximal  supergravity soft limits}
The supercharges are symplectic Majorana-Weyl spinors with positive 
chirality $Q_+^i$ ($i = 1, 2, \cdots, N_+$) and symplectic 
Majorana-Weyl spinors with negative chirality $Q_-^i$ 
($i = 1, 2, \cdots, N_-$). They satisfy 
$\Omega_+^{ij} (Q_+^j)^c = Q_+^i$, 
$\Omega_-^{ij} (Q_-^j)^c = Q_-^i$, 
where $\Omega_\pm^{ij}$ are antisymmetric matrices. 
The numbers $N_+$ and $N_-$ must be even. 
The automorphism group is  USp($N_+$) $\times$ USp($N_-$). 
Anticommutators of the supercharges are 
\ba
\{Q_+^i, Q_+^{jT} \} \A \ = \  \A {1 \over 2} 
(1+\bar\gamma) \gamma^a C_- P_a \Omega_+^{ij} \ , \nonu
\{Q_-^i, Q_-^{jT} \} \A = \A {1 \over 2} 
(1-\bar\gamma) \gamma^a C_- P_a \Omega_-^{ij} \ , \nonu
\{Q_+^i, Q_-^{jT} \} \A = \A {1 \over 2} 
(1+\bar\gamma) C_- Z^{ij} \ . 
\ea
Consider maximal 6D supergravity and 1/2 BPS superfields.
\be
N_+=4\, , \qquad  N_-=4 \ .
\ee
The R symmetry group is  $USp(4) \times USp(4)$. It coincides with the isotropy group
${\cH}$ associated with the coset ${\cG\over \cH}= {E_{5(5)}\over USp(4) \times USp(4)}$. It is convenient to discuss the ultrashort multiplets using the following choice of the 4x4 
$\Omega_{\pm}$-matrix \cite{Ferrara:2000xg}
\be
\left(\begin{array}{cc}0 & {\cal I}  \\-{\cal I}  & \, 0\end{array}\right) \, ,
\label{Omega}\ee
here ${\cal I}$ is the 2x2 identity matrix. The non-vanishing entries in the RHS of supersymmetry algebra are 
\be
\Omega^{13}_{\pm} = \Omega^{24}_{\pm} =1 \ .
\ee
Spinor coordinates of the superspace satisfy a Majorana-Weyl pseudoreality condition
\be
\bar \theta^i =\Omega^{ij} \theta_j^\beta c_{\beta \alpha} \ .
\ee
The spinor derivatives satisfy the supersymmetry algebra 
\be
\{D_\a^i, D_\beta ^j\}= i \Omega^{ij} \gamma^\mu \partial_\mu \ .
\ee
The 25 scalars  are given by $V_a^{\hat a}$ field in $SO5)\times SO(5)$ form, $a, \hat a=1,2,3,4,5$. The BPS superfield is described  in $USp(4) \times USp(4)$ basis so that the first component is
\be
V_{ij}^{ \hat i \hat j}= \gamma^{a}_{ ij} V_a^{\hat a} \gamma_{\hat a}^{ \hat i \hat j} \ .
\ee
Using linearized supersymmetry of the 6D maximal supergravity action \cite{Tanii:1984zk,Bergshoeff:2007ef} one finds that the superfield $W_{12}^{\hat 1 \hat 2}(x, \theta)$ depends only on half of fermionic coordinates
\be
D_{\a 1} W_{12}^{\hat 1 \hat 2} = D_{\a 2} W_{12}^{\hat 1 \hat 2} = D^{\hat \a \hat 1} W_{12}^{\hat 1 \hat 2} D^{\hat \a \hat 2} W_{12}^{\hat 1 \hat 2}=0   \ .
\label{12BPS}\ee 
This is consistent with the choice of the symplectic matrix $\Omega$ in \rf{Omega}. By making this choice of the superfield, we have broken $USp(4) \times USp(4)$ R symmetry down to its subgroup $SU(2)\times SU(2)$.

Now we can apply this information to 6D multi-point superamplitudes along the lines used for type IIB supergravity amplitudes in 10D in \cite{Wang:2015jna}.
The $n$-point amplitude is defined as 
\be
{\cal A}=\delta^6(P)\delta^{16}(Q) {\cal F} (\lambda_i, \eta_i) \ .
\ee
Here
\be \label{eq:sugra-4pts}
\delta^{16}(Q)=  \delta^{8} \left( \sum_{i=1}^n q^{A, I}_i \right)  \delta^{8} \left( \sum_{i=1}^n {\hat q}^{ {\hat I} }_{i,{\hat A}}\right)  \, ,
\ee
which has manifest permutation symmetry. Here the supercharges are defined as $q^{A, I}_i = \lambda^A_{i, a} \eta^{I, a}_i$, and $  {\hat q}^{ {\hat I} }_{i,{\hat A}}= \hat{\lambda}_{i, \hat{A},\hat{a} } \hat{\eta}^{ {\hat I}, \hat {a}}_i$. These are half of the supercharges, and the other half involve $\eta$ derivatives. $I, \hat I=1,2$ and $ A, \hat A=1,2,3,4.$ Conservation of these additional supercharges automatically follows from the first set together with the R symmetry.
With 
\be
Q = \sum_{i=1}^n \lambda_i \, \eta_i  \qquad \tilde Q = \sum_{i=1}^n \lambda_i \, {\partial \over \partial \eta_i } \ , 
\ee
we recover the susy algebra in \rf{susy}.
The amplitude is manifestly supersymmetric under the action of the $Q$ operator, i. e. under 1/2 of supersymmetry. The action of $\tilde Q$ is more constraining, in general. We have already explained this in \cite{Kallosh:2023dpr} in terms of superinvariants. At that time, the relevant information from the point of view of superamplitudes was proposed in the context of 10D type IIB supergravity amplitudes. Here we can do it directly in the framework of linearized 6D maximal supergravity amplitudes.

Consider R-invariant multi-point superamplitudes in 6D.
The $n$-point superamplitudes with $n>4$ at $L$-loop order satisfying linearized Ward Identity at maximal supergravity are
\be
 \kappa^{2(L-1)}\delta^6(P)\delta^{2\cN}(Q) \tilde Q \, h(\lambda_i, \eta_i) \ ,
\label{6dn}\ee
where according to \rf{nBPSc} in 6D and $\cN=8$
\be
{\rm dim} [ h(\lambda_i, \eta_i)]= 4 L - 14 \ .
\label{nBPS6d}\ee
Thus, at $L=3$ we find that 
\be
{\rm dim} [ h(\lambda_i, \eta_i)]= 12 - 14=-2 \ .
\label{nBPS6dL3}\ee
and at 3-loops, there are no supervertices! There was again no need actually to find the explicit solutions of  Ward identities for the 6-point amplitudes, as we did in \cite{Freedman:2018mrv} in 4D case. It was sufficient to establish the dimension of the function $h(\lambda_i, \eta_i)$, as we did also above for the 4D $\cN\geq 5$ supergravities in Sec. \ref{Sec:4d}.

 This, in turn, means that the 4-point UV divergence at $L=3$ maximal 6D supergravity breaks $E_{5 (5)}$  symmetry since the soft scalar limit of the total 6-point amplitude is broken.
 
 Thus, in addition to the analysis of superinvariants, we have also shown that the superamplitudes in 6d supergravity lead to the same conclusion.
 
 A nice example of the 10D 6-point superamplitude was already given in \cite{Green:2019rhz} and discussed in  \cite{Kallosh:2023dpr}. In 6D, the 3-loop UV divergence is of the type $\kappa^{4} \int d^4 x \partial^6 R^4$. We are looking at the local $SO(5)\times SO(5)$ invariant amplitude with two additional scalars
 \be
 \kappa^{4} \int d^4 x \partial^6 R^4 \phi_a^{\hat a} \phi_b^{\hat b} \delta^{ab} \delta_{\hat a \hat b}\, , \qquad a, \hat a=1,2,3,4,5 \ .
 \ee
A supersymmetric one would take a form  \rf{nBPS}
\be
\kappa^{4} {\cal A}_{non-BPS} = \kappa^{4} \delta^6(P)\delta^{16}(Q) \tilde Q \, h(\lambda_i, \eta_i) \ .
\label{nBPSaa}\ee
We already know that dimension of $h$ here is negative, so the superamplitude has a pole. It is interesting to see how it depends on $\lambda$ and $\eta$'s. 

The scalar superfields have a graviton at the 4th component, so to have 4 gravitons we need 16 $\eta$'s. One of the scalars has $\eta^0$, the other one has $\eta^8$, one is the first component of the BPS superfield, the other one is the last. So we need the total of  $\eta$'s in $h$ to be 24. 
 This is literally the superamplitude given in \cite{Green:2019rhz} for 10d type IIB supergravity and discussed in \cite{Kallosh:2023dpr}
\be
 \delta^{16} (Q_n) (\bar Q_n)^{16} (\eta_i)^8 (\eta_j)^8 (\eta_k)^8 \ ,
\label{wen}\ee
To be more specific, in 6D at 3-loop order, we have the following superamplitude, which includes  4 gravitons and 2 scalars which is $USp(4) \times USp(4)$ invariant
\be
\kappa^4 \delta^6(P) \delta^{16} (Q_n) (\bar Q_n)^{16} (\eta)^{24}  h(\lambda)\ , \qquad {\rm dim} \, h(\lambda) = -2 \ .
\label{us}\ee
Note that we have 32 $\lambda$'s in $\delta^{16} (Q_n) (\bar Q_n)^{16}$ which gives 16 in momenta,  and only -14 from $\kappa^4$ and $\delta^6(P)$ at $L=3$.
Now we can see why the 10D superamplitude in \rf{wen} is in agreement with our superinvariant analysis in 6D in \cite{Kallosh:2023dpr} as well as our new analysis here of the superamplitudes satisfying Ward identities in 6D as given in eq. \rf{us} here.

The relevant 6-point candidate 3-loop linearized superinvariants does not exists since the  action  has a negative dimension
\be
{\rm dim} \Big [ \kappa^4  \int d^6 x \,  d^{32} \theta \,  Tr \, V^6 \Big ] =-2 \ ,
\label{L36p}  \ee
Here 
\be
Tr \, V^6= V_a^{\hat a} V_b^{\hat b} V_c^{\hat c} V_d^{\hat d} V_e^{\hat e} V_f^{\hat f} t^{abcdef}_{\hat a \hat b \hat c \hat d \hat e\hat f}
\ee
and $t^{abcdef}_{\hat a \hat b \hat c \hat d \hat e\hat f}$ is an $SO(5)\times SO(5) $ covariant tensor. $V_a^{\hat a}$ is a scalar superfield starting with a 25-components scalar field $\phi_a^{\hat a}$.
Its dimension is the same as in a relevant superamplitude in eq. \rf{us}.

\section{Summary}
Supersymmetric Ward identities for linearized extended  $\cN\geq 5$ supergravities
 in 4D are known for a  long time \cite{Bianchi:2008pu,Elvang:2009wd,Freedman:2018mrv}. 
 We have found here that the more recent understanding of supersymmetric Ward identities in 10D type IIB supergravity in \cite{Wang:2015jna,Green:2019rhz} allows a nice and useful generalization of these identities to  $D\geq 4$ and extended  $\cN\geq 5$ supergravities.

It is satisfying to see this kind of universality with regard to Ward identities for amplitudes. The short summary is 
that  

1. 1/2 BPS  superamplitudes have the form
\be
 \kappa^{2(L-1)} \delta^D(P)\delta^{2\cN}(Q) f(s_{ij}) \ .
\ee
 and 
\be
{\rm dim}[ f(s_{ij})]= (D-2) L +2-\cN \ .
\ee
This is to be compared with the 1/2 BPS linearized superinvariants of the form where we integrate over 1/2 of the superspace a Lagrangian depending on superfields 
\be
 \kappa^{2(L-1)} \int d^D x d^{2\cN}\theta \, {\cal L}^{BPS} \Big (\phi(x, \theta)\Big) \ ,
\ee
and 
\be
{\rm dim} [ {\cal L}^{BPS} (\phi(x, \theta))]= (D-2) L +2- \cN \ .
\label{nBPScDis}\ee
Thus 
\be\boxed{
{\rm dim} [ {\cal L}^{BPS} (\phi(x, \theta))]= {\rm dim}[ f(s_{ij})]}
\ee

2.  non-BPS $n$-point R-invariant superamplitudes with $n>4$ have the form
\be
 \kappa^{2(L-1)} \delta^d(P)\delta^{2\cN}(Q) \tilde Q \, h(\lambda_i, \eta_i) \ ,
\label{nBPSdis}\ee
 and 
\be
{\rm dim} [ h(\lambda_i, \eta_i)]= (D-2) L +2- 2\cN \ .
\label{nBPScDis}\ee
This is to be compared with the non-BPS linearized superinvariants of the form where we integrate over the whole superspace a Lagrangian depending on superfields 
\be
 \kappa^{2(L-1)} \int d^D x d^{4\cN}\theta \, {\cal L}^{non-BPS} \Big (\phi(x, \theta)\Big) \ ,
\ee
and 
\be
{\rm dim} [ {\cal L}^{non-BPS} (\phi(x, \theta))]= (D-2) L +2- 2\cN
\label{nBPScDis}\ee
Here we also note that
\be\boxed{
{\rm dim} [ h(\lambda_i, \eta_i)]= {\rm dim} [ {\cal L}^{non-BPS} (\phi(x, \theta))]} \ .
\ee
If the dimension of ${\cal L} (\phi(x, \theta)$ is negative, the superinvariant is not a local one; some insertion of inverse derivatives is required like $\partial^{-2}$. But if the dimension of ${\cal L} (\phi(x, \theta)$ is not negative, we have a local superinvariant. If the dimension of the function $h(\lambda_i, \eta_i)$ is negative, it means that  $\lambda$'s are in denominator, and there are some poles in momentum space since $p$ is quadratic in $\lambda$'s. Here $\eta$'s are dimensionless. If the dimension of $h(\lambda_i, \eta_i)$ is non-negative, the corresponding superamplitudes are so-called supervertices; they have no poles in momentum space. It is, therefore, not surprising that superamplitudes and linearized superinvariants have the same predictions.

The examples are in 4D $\cN\geq 5$, where there are no 6-point supervertices at $L < \cN-1$,
\be
{\rm dim} [ h(\lambda_i, \eta_i)]= {\rm dim} [ {\cal L}^{non-BPS} (\phi(x, \theta))]= 2 (L -(\cN-1)) \ ,
\ee
since at $L < \cN-1$, both superamplitudes as well as superinvariants are non-local. In  \cite{Beisert:2010jx}  and in \cite{Freedman:2018mrv} it was shown using superamplitudes as well as superinvariants.

In maximal 6D 
\be
{\rm dim} [ {\cal L}^{non-BPS} (\phi(x, \theta))]= {\rm dim} [ h(\lambda_i, \eta_i)]=4 L -14 \ .
\label{nBPSc6d}\ee
The absence of 6-point supervertices at $L\leq 3$ was proven in \cite{Kallosh:2023dpr}  using linearized superinvariants. Here we have shown that the same result follows from superamplitudes satisfying Ward identities.

The nice thing we have learned here is that it was not necessary to compute explicitly the 6-point amplitudes satisfying Ward identities, as we did in 4D case in \cite{Freedman:2018mrv} for $L\leq \cN-1$. It was sufficient both in 4D and in 6D cases to establish a dimension of the function $h(\lambda, \eta)$
which appears in the solutions of Ward identities. That allowed us to find out here that certain 6-point amplitudes are not local, which was important for the soft scalar limits studies.

The actual computation of the $n$-point amplitudes is necessary when we are looking for the 1-loop anomalies. This was the case in 4D $\cN=4$ supergravity  \cite{Carrasco:2013ypa}. The existence of linearized $n$-point half-BPS superinvariants with non-vanishing soft scalar limits was established, for example, there was a term
\be
{d_{n+2}\over (4\pi)^2} \int d^4 x d^8\theta  W^{n+2} +cc
\ee
where the actual value of $d_n$ was not known. The corresponding superamplitude with 2 gravitons and $n$ scalars was also known to be proportional to 
$
 \delta^8 \Big (\sum_{i=1}^n \tilde \eta_i \tilde \lambda_i \Big)
$.
But the actual computation of this amplitude has confirmed that 
$
d_{n+2}= {1\over 2(4\pi)^2} {1\over n(n+1) (n+2)} \neq 0
$
and that the 1-loop anomaly is present.

There was another candidate for 1-loop anomaly $U(1)$ anomaly in 4D $\cN=4$ supergravity in the form of a linearized superinvariant 
\be
f_4 \int d^8\bar \theta {1\over \partial^6}  C \, \partial \partial \overline W \, \partial \partial \overline W \,  \overline W
\ee
The superamplitude matching this superinvariant was computed in \cite{Carrasco:2013ypa} using the double-copy method. The result has shown that $f_4\neq 0$ and the corresponding 1-loop anomaly is present.

The fact that some superamplitudes are consistent with Ward identities or, equivalently, can be presented as superspace integrals can only tell us that such amplitudes are consistent with linearized supersymmetry. But only the actual computation can tell us if these are present or not. For example, the 1-loop 6-point amplitudes in 6D maximal supergravity have to be computed to find out if there are $E_{5(5)}$ 1-loop anomalies.

\

\noindent{\bf {Acknowledgments:}} I am grateful to J. J. Carrasco, H. Elvang, D. Freedman, Y.-t. Huang,  A. Linde, J. Parra-Martinez, R. Roiban, C. Wen, and  Y. Yamada for stimulating discussions of the Ward identities and of the soft scalar limits in supergravities. 
 This work is supported by SITP and by the US National Science Foundation grant PHY-2310429.  

 \appendix
 \section{One-quarter and  one-eighth  BPS superinvariants}\label{App:A}
 In this paper, we have studied only half of BPS and non-BPS superinvariants and the relevant superamplitudes. On the other hand, there are examples of linearized superinvariants, which are one-half, one-quarter, and one-eighth superinvariants \cite{Bossard:2009sy}. These are based on superfields in harmonic superspace, which depend on  $4\cN(1-{1\over k})$ directions, with $k=2,4,8$. Ignoring harmonic coordinates, they are given as integrals over a subspace of the whole superspace of the form
 \be
 \int d^D x d^{4(\cN-{1\over k})} \theta {\cal L} (\phi(x, \theta))
 \ee
 where the superfields $\phi(x, \theta)$ depend only on a fraction of all $\theta$-directions. For example, in maximal $\cN=8$ case, one-half of BPS superfields depend on 16 $\theta$'s, one-quarter of BPS superfields depend on 24 $\theta$'s, and one-eighth  BPS superfields depend on 28 $\theta$'s.
 
 All 1/k BPS superinvariants have R-symmetry for 4-point superinvariants. However,  when the number of superfields exceeds 4 in all examples, the R-symmetry of such invariants is broken.
 
In the context of superamplitudes, which we studied in this paper, the cases of one-half-BPS and non-BPS are simple and are now fully described in this paper for dimensions $D\geq 4$ and  $\cN\geq 5$. 
 
We do not know of examples of one-quarter and one-eighth  BPS superamplitudes corresponding to relevant known superinvariants,  and whether these superamplitudes are available or not. Nevertheless, we brought it up here to keep in mind that there are cases with linearized superinvariants that have not been studied in the context of superamplitudes.

\bibliographystyle{JHEP}
\bibliography{refs}
\end{document}